\documentclass[review]{elsarticle}

\usepackage{hyperref}
\usepackage{float}
\usepackage{placeins}
\usepackage{mathtools}
\usepackage{amsfonts}
\usepackage{amsmath}
\usepackage{amssymb}
\usepackage{graphicx}
\usepackage{float}
\usepackage{verbatim}
\usepackage{array}
\usepackage{multirow}
\usepackage{dcolumn}
\usepackage{color}

\usepackage{setspace}
\usepackage{diagbox}
\usepackage{adjustbox} 
\usepackage{makecell}
\usepackage{mathtools}
\usepackage{comment}

\usepackage{geometry}
\journal{Journal of \LaTeX\ Templates}

\biboptions{authoryear}

\begin{document}
%\title{On the Periodic Autoregressive Model Applied to Inflow Forecast on the Brazilian Energy System}

\begin{frontmatter}

\title{Assessing the Optimistic Bias in the Natural Inflow Forecasts: A Call for Model Monitoring in Brazil}

\author{Arthur Brigatto\textsuperscript{1},
Alexandre Street\textsuperscript{1}*,
Cristiano Fernandes\textsuperscript{1}, 
Davi Valladão\textsuperscript{2},
Guilherme Bodin\textsuperscript{3}
and
Joaquim Dias Garcia\textsuperscript{3}
}
\address{\textsuperscript{1}Electrical engineering Department at the Pontifical Catholic University of Rio de Janeiro (PUC-Rio). \\ \textsuperscript{2}Industrial engineering Department at the Pontifical Catholic University of Rio de Janeiro (PUC-Rio). \\ \textsuperscript{3}PSR, Rio de Janeiro.}

%\author[mymainaddress]{Pontifical Catholic University of Rio de Janeiro}

\cortext[mycorrespondingauthor]{Corresponding author: Alexandre Street (street@puc-rio.br)}
%\ead{street@puc-rio.br}


\begin{abstract}
    
Hydroelectricity accounted for roughly {61.4\% of Brazil's total generation in 2024} and addressed most of the intermittency of wind and solar generation. Thus, inflow forecasting plays a critical role in the operation, planning, and market in this country, as well as in any other hydro-dependent power system. These forecasts influence generation schedules, reservoir management, and market pricing, shaping the dynamics of the entire electricity sector. { The objective of this paper is to measure and present empirical evidence of a systematic optimistic bias in the official inflow forecast methodology, which is based on the PAR(p)-A model. Additionally, we discuss possible sources of this bias and recommend ways to mitigate it.}
By analyzing 14 years of historical data from the Brazilian system through rolling-window multistep (out-of-sample) forecasts, results indicate that the official forecast model exhibits statistically significant biases of 1.28, 3.83, 5.39, and 6.73 average GW for 1-, 6-, 12-, and 24-step-ahead forecasts in the Southeast subsystem, and 0.54, 1.66, 2.32, and 3.17 average GW in the Northeast subsystem. These findings uncover the limitations of current inflow forecasting methodologies used in Brazil and call for new governance and monitoring policies.

\end{abstract}

\begin{keyword}
Brazilian power system \sep forecast bias \sep hydrothermal power system \sep inflow forecast \sep operation planning \sep periodic autoregressive \sep renewable energy.
%\MSC[2010] 00-01\sep  99-00
\end{keyword}

\end{frontmatter}

% Use this to place sponsorships

\section{Introduction} \label{Introduction}
From short-term generation scheduling and market pricing to long-term planning, models and their data play a central role in power systems worldwide. Policymakers can pass new laws and design new markets, but if the new guidelines are not well reflected, if data is imprecise, or if the forecasts feeding the chain of models used to dispatch, plan, or price electricity are biased, the functioning of the entire system and market can be at risk. In this paper, we present concerning evidence that systematically biased inflow forecasts are being used to feed the entire chain of models in Brazil. In the following, we provide a comprehensive motivation for the topic, highlight the regulatory willingness for changes in Brazil, and define the aim and objective of the paper.

With an annual consumption of approximately 80 average GW (700 TWh) in 2024, hydropower is the primary renewable source for electricity generation within the Brazilian energy infrastructure, accounting for 61.4\% of total generation. It is followed by intermittent renewables (25.9\%, with 15.2\% from wind and 10.7\% from solar) and thermal sources (12.8\%) \citep{ONS2025}. Even in cases of drought, hydroelectric plants still account for the majority of electricity generation in this country. In this context, a long-term multistage dispatch planning model, generally based on the Stochastic Dual Dynamic Programming technique (SDDP, see \cite{pereira1991multi}), is used to estimate the opportunity cost of water, which is then used in the short-term operation procedures to define the day-ahead scheduling and market prices. However, this multistage operational policy heavily relies on multiperiod forecasts (e.g., 60 months in Brazil) of water inflow time series. Hence, any systematic bias in the inflow forecasts can impact the water values and compromise the optimal usage of this scarce resource over time. In hydro-dependent power systems, such as Chile, Colombia, Mexico, Norway, and Vietnam, to mention a few, this process varies from country to country, depending on the market design. Still, its heavy reliance on inflow forecasts is typically unchanged.

In Brazil, the official implementation of the SDDP model utilizes aggregated reservoirs, combining the total natural water inflow within each subsystem, which is frequently expressed in terms of energy values for long-term planning. This process results in what is known as the Natural Inflow Energy (NIE) time series. Natural inflow time series have been widely studied in academic papers in various contexts (see \cite{paula2015}, \cite{erlon2024}, and references therein). Interestingly, the NIE serves as an important planning metric, representing the additional energy that the system (or subsystems) receives from river inflows. So, because the Brazilian system still relies on more than 60\% of hydro generation, any issue associated with the forecasts of such a relevant energetic resource in Brazil is of high concern.

The literature on inflow forecasting is vast. {For instance, in \cite{farfan2024streamflow}, a representative hybrid deep learning inflow forecast methodology was used to forecast daily inflow in Turkey. Four different architectures were tested and benchmarked to forecast the next day based on external variables associated with meteorological data. Within AI-based models, machine learning techniques applied to inflow forecast also constitute a relevant research topic (see \cite{MLreview} for a recent review on the topic). Interesting studies on how relevant climate phenomena, such as El Nino and La Nina, affect inflow patterns \citep{paula2019} and influence Latin American markets \citep{Colombia} provide us with relevant examples on the complexity of mid-term forecasts and their interdependence with hard-to-predict climate variables.}

{Interestingly, a language modeling framework adapted for time series forecasting and trained with a large set of real and synthetic data \citep{chronos}, has proven to outperform classical and deep learning methods. As per \cite{erlon2024}, Temporal Fusion Transformers (TFT \citep{TFT}) provide interesting gains from the deep learning side for short-term inflow forecasts. So, machine learning and deep learning have been demonstrated to be an interesting tool for combining with classical methods to improve short-term inflow forecasting (see \cite{erlon2024}, \cite{MLreview}, and \cite{zhao2024review} and references therein). } 

{However, to comply with the SDDP convexity hypothesis, previously reported nonlinear deep learning or machine learning forecasting models cannot be used directly. Therefore, for mid- and long-term operational planning purposes, inflow time series in Brazil—and in most countries using SDDP-based planning tools—have been forecasted using linear Periodic Autoregressive (PARp) models, an approach that has remained largely unchanged for more than three decades. We refer the interested reader to \cite{pereira1984stochastic} and \cite{noakes1985forecasting} for pioneer publications, and to \cite{MATOS20121443}, \cite{shapiro2013risk}, and \cite{GTMetodologiaCPAMP2022} for more recent advances. For instance, in \cite{shapiro2013risk} an interesting yet limited multiplicative version of the PARp model was studied, and more recently, the official Brazilian PARp model was replaced by the PARp-A model (see \cite{GTMetodologiaCPAMP2022}), which adds an annual (12-month) moving-average term to the original PARp model. The PARp-A was motivated by evidence that the PARp model was unable to capture longer inflow profiles below the long-term mean (LTM) observed in the historical data.} 

Within the context of optimal dynamic-programming-based policies \citep{pereira1991multi}, an unbiased assessment of the cost-to-go function ---which comprises the optimal operating cost from the second stage onward--- is crucial for achieving the optimal balance between immediate (first-stage decisions) and future dispatch costs, and thereby driving the system towards a safe and cost-effective storage path. In contrast, optimistically biased NIE forecasts, especially for $k\ge2$ steps ahead, can potentially introduce bias in the dispatch decisions and lead the system to riskier storage levels. Therefore, it is crucial to monitor the multiple-step-ahead NIE forecast bias to avoid biasing the water values and, consequently, generation scheduling decisions and prices.

From the empirical side, serious droughts have been observed in Brazil since 2013, with the most recent one occurring in the second semester of 2021 (see \cite{w14040601}). However, the dispatch orders suggested by the official models often delayed the activation of preventive actions, such as thermoelectric dispatches in pre-crisis situations --- even though it was clear that these actions were needed at that time. This recurring pattern has led to frequent out-of-merit dispatches, with high costs ultimately passed on to consumers. In 2021 alone, these costs reached more than R\$19 billion\footnote{about 3.8 billion USD on the mean currency exchange value in 2021} (see Figure 33 in \cite{ONS_out_of_merit} for an official evaluation of this cost by the Brazilian System Operator). 

Interestingly, at the same time, observed inflows have been systematically falling short of forecasts in the last decade. Thus, as the water stored in hydropower reservoirs constitutes the most relevant generation resource in the Brazilian power system, the question of whether optimistic forecasts contributed to the delayed activation of preventive resources in pre-crisis situations raised awareness of most entities and agents in the sector. So, the first step in investigating the impact of optimistic inflow forecasts on the Brazilian power system is to provide empirical evidence of its significance and magnitude.

Exploring the academic side, there is reasonable evidence reported in the literature regarding the detrimental effects of optimistic bias in the water value assessment due to simplifications in the operational models (see \cite{brigatto2017assessing, street2020assessing}, and \cite{rosemberg2021assessing}). Other Latin American countries also identify the need for monitoring the optimistic bias in the water value assessment (see 4.3 in \cite{ISCI2024}). Therefore, based on the previously reported empirical and academic evidence, since 2022, the Brazilian electricity sector has begun to discuss the topic more seriously. The regulatory willingness to change the governance of models and data triggered a series of public calls for contributions, which led to the National Council for Energy Policy approving Resolution No. 1, dated March 12, 2024 (officially published on April 19, 2024). This resolution established guidelines aiming to ensure the coherence and integration of input data, parameters, methodologies, and computational models utilized by the Ministry of Mines and Energy, the Energy Research Company (the planner), the National System Operator, and the Electric Energy Trading Chamber (the market operator)\footnote{In \cite{street2024modelgovernance}, a series of papers and articles depicting the problem is provided. In Occurrences, the main official decisions are provided. The aforementioned resolution is translated and made available in English.}.

Recognizing that it is in the best interest of system operators to monitor and identify any source of unintentional bias in official models, \emph{this paper aims to contribute to current industry practices and regulatory policies by presenting empirical evidence of the forecasting bias in Brazil’s official PARp-A model {and calling for an improved model and data governance framework}. %this paper aims to contribute to current industry practices and regulatory policies by measuring and presenting empirical evidence of the forecasting bias in Brazil's official PARp-A model and calling for a new model and data governance. 
Additionally, we discuss this bias and recommend ways to mitigate it.} This is achieved by measuring positive first-moment shifts (optimistic forecast biases) in the forecasted distributions relative to the observed data. Acknowledging that the official dispatch models use multi-step-ahead forecasts to assess the opportunity cost of storing the water for future stages, we evaluate the bias of the official NIE forecasts for 1 to 24 months ahead. In addition, as the cost-to-go function estimation relies on the weighted sum of future costs, considering the system's capacity to transfer water from one stage to another, which is typical of the high storage capacity in Brazil, the bias in the cumulative NIE forecasts is also evaluated and analyzed. 
%To properly assess the bias of the current official forecasting model, acknowledging the fact that the official water-value calculation methodology uses multi-step-ahead forecasts, bias metrics for 1 to 24 months ahead. Results for both the individual and cumulative forecasts are provided and studied. 

By analyzing 14 years of historical data through rolling-window (out-of-sample) multistep forecasts, our results indicate that the official forecast model exhibited statistically significant biases in the Southeast and Northeast subsystems. Based on these findings, this paper raises awareness of the magnitude of the potential bias that may be introduced into procedures that currently rely on these forecasts, including generation scheduling, short-term pricing, and expansion planning. Additionally, the identification of a decade-long bias should also motivate local authorities to implement new monitoring procedures based on specialized, application-specific metrics to detect and control the bias in inflow forecasts with greater agility than has been previously achieved. 

Although the main contribution of this paper lies in the governance of forecasting models, specifically in estimating and reporting the inflow forecast bias in Brazil, and discussing how it could be monitored, we also address several interesting and relevant methodological challenges regarding the metric and assessment procedure. Therefore, the rolling-window evaluation scheme used to assess and monitor the out-of-sample bias of forecasting models is provided through an open-source Julia package \citep{CompareScenariosGenerators}. Finally, the NIE data time series used in this paper and forecast results are all available in \cite{CaseStudyData}.   

%Finally, a comprehensive description of the estimation procedure of the PARp and PARp-A models can be found in various academic papers (\cite{pereira1984stochastic}, \cite{noakes1985forecasting},  and \cite{GTMetodologiaCPAMP2022}). However, to the best of the authors' knowledge, there is no previous scientific work in which the simulation procedures and forecasts of the PARp and the PARp-A models are clearly presented. Thus, as these details are relevant for reproducibility purposes, we provide them in our Appendix. 

%Finally, it is relevant to mention that this paper does not aim to solve the bias issue. It rather aims to demonstrate, through an objective methodology, the existence of such a significant and arguably high gap between what is expected from the theory and hypothesis of the adopted models and the practice. We hope that this work will be helpful for both academics and practitioners in 1) further improving the official Brazilian NIE forecasting methodology and 2) establishing new monitoring procedures to accelerate correction actions.   

\vspace{-0.2cm}
\section{Forecast Bias Assessment} \label{metrics}
%the ongoing process of tracking and evaluating the accuracy and performance of predictions or forecasts over time.

This section presents a framework for evaluating and monitoring the accuracy of NIE predictions made by the PARp and PARp-A models. As the introduction states, the official operational methodology uses a multistage problem to assess the water values to dispatch units in the short term. The water values are calculated as the derivative of the present value of the optimal operating costs over the 60 months considered in the planning horizon. Hence, it is important to monitor the forecasts for several steps ahead ($k=1,...,K$) as they will all build up the water value. Although crucial for this application, this is not standard in the statistical literature, which mainly relies on one-step-ahead assessments. 

\subsection{Natural Inflow Energy (NIE) in Brazil}

Due to historical reasons and specific modeling choices, the official SDDP implementation in Brazil continues to rely on aggregated reservoir management. In this approach, the total natural water inflow within a given subsystem is converted into the subsystem's equivalent NIE. Essentially, each subsystem's NIE time series represents the total energy that would have been produced if all the water inflow at various points along the cascade had flowed through the corresponding downstream hydro turbines to the river's endpoint. This procedure computes the total generated potential of water inflows in each subsystem and is widely used in the operational planning of hydrothermal systems \citep{MATOS20121443}, \citep{street2020assessing}. For a detailed explanation of the NIE concept and the aggregation method, we refer to \cite{de2008comparison}. For reproducibility purposes, the historical data used in this paper and its results are available at \cite{CaseStudyData}.

\subsection{Error and Bias Metrics}

In this context, we propose a bias monitoring procedure based on a rolling-window scheme to evaluate the out-of-sample $k$-steps-ahead forecast errors and assess the model bias. In this procedure, for each period $t$ in a subset of the historical records, the forecast model is estimated with data up to period $t$ (represented by $y_{[t]}$, a vector with all observed data up to $t$) and a forecast $\hat{y}_{t+k|y_{[t]}}$ for $k=1,...,K$ steps ahead is obtained.\footnote{Data leakage is a risk in this step. So, the best practices in data manipulation should be adopted to avoid using data beyond $t$ in the estimation process.} Then, a $K$-dimensional vector of forecast errors for $t$ is calculated by contrasting the forecast with the previously unseen data, now revealed to the evaluation process, as $\hat{\varepsilon}_{t,k} = \hat{y}_{t+k|y_{[t]}} - y_{t+k}$. This process is repeated following a rolling-window scheme for a given subset of period $t$ of the historical data. The collected error vectors are then used to evaluate the bias and error metrics. For the sake of conciseness and didactic purposes, we will present the metrics focusing on only one time series, assuming the forecast procedures are given. Finally, for reproducibility purposes, we provide in \cite{CompareScenariosGenerators} an open-source package that was used to evaluate all metrics presented in this work.

We start by defining the bias of the $k$-steps ahead forecast as follows:
\begin{align}
\Delta_{k}^{(\overline{MW})} = \dfrac{1}{n_k}\sum_{t=1}^{n_k}\hat{\varepsilon}_{t,k},\label{bias_MWm}
\end{align}
%where $\hat{y}^{(Mean)}_{t+k|y_{[t]}}$ is the average across all the scenarios in $\Omega$ simulated (by Monte Carlo procedure) for month $t+k$ from month $t$, $y_{[t]}$ is a vector of observed data up to period $t$, $y_{t+k}$ is the observed NIE in month $t+k$, and 
where $n_k$ is the number of out-of-sample $k$-steps-ahead forecasts. 
The 95\% confidence interval for the bias is calculated as follows \citep{brockwell2002introduction}:
\begin{align}
\Bigg(\Delta_{k}^{(\overline{MW})} - 1.96\sqrt{\dfrac{\hat{v}_{k}}{n_k}}, ~\Delta_{k}^{(\overline{MW})} + 1.96\sqrt{\dfrac{\hat{v}_{k}}{n_k}} \Bigg),\label{CI}
\end{align}
where $\hat{v}_{k} = \sum_{|h|<\sqrt{n_k}}\bigg(1-\dfrac{|h|}{\sqrt{n_k}}\bigg)\hat{\gamma}(h)$ and $\hat{\gamma}(h)$ is the sample autocovariance function. To obtain $\hat{\gamma}(h)$ we firstly evaluate $\hat{\varepsilon}_{t,k}$ for a given $k$ and all $t\in \{1,\dots,n_k\}$ to estimate a time series of $k$-steps-ahead forecast errors. Then, the autocovariance is computed.

\section{The PARp and PARp-A Models}
\label{Parp:section}
 In this section, we outline the official implementation (estimation and simulation) of the PARp and PARp-A models for the NIE used in the Brazilian model. %While certain aspects of this implementation may be documented in official technical notes and other literature, a comprehensive exposition of the implementation is not found in any single publication. 
 We start by defining a monthly time series $\{y_{t}\}_{t=1}^T$ of NIE for a given subsystem. We use $m \in M = \{1,...,12\}$ for months and $m_t$ to denote the month in $M$ of time $t$. The estimates for the historical LTM and standard deviation of each month are defined as $\hat{\mu}_{m}$ and $\hat{\sigma}_{m}$, respectively, for $m \in M$. Following \cite{noakes1985forecasting}, the PARp model is defined through a set of 12 normalized autoregressive models, each of which with parameters $\phi_1^{(m)},...,\phi_{p_m}^{(m)}$, as follows: 
\begin{align}
    &\bigg(\dfrac{y_{t} - \hat{\mu}_{m_{t}}}{\hat{\sigma}_{m_{t}}}\bigg) = \phi_{1}^{(m_{t})}\bigg(\dfrac{y_{t-1} - \hat{\mu}_{m_{t-1}}}{\hat{\sigma}_{m_{t-1}}}\bigg) + \dots + \nonumber  \\ &+\phi_{p_{m_{t}}}^{(m_{t})}\bigg(\dfrac{y_{t-p_{m_{t}}} - \hat{\mu}_{m_{t}-p_{m_{t}}}}{\hat{\sigma}_{m_{t}-p_{m_{t}}}}\bigg) + \epsilon_{t}.\label{PARp_model}
\end{align}
In \eqref{PARp_model}, $\epsilon_{t}$ is the random error term of the normalized time series, having zero mean and periodic standard deviation $\sigma^{(\epsilon)}_{m_t}$, and $p_{m_t}$ is the order of the autoregressive model, both for month $m_t$ for each $t$. %Following the estimation process, the estimated standard deviations, denoted as $\hat{\sigma}^{(\epsilon)}_{m}$, can be computed from the data for each month $m \in M$.

The formulation of the PARp-A model is very similar to that of the PARp model, with the sole distinction being the inclusion of a fixed term representing the average of the last $12$ months. Specifically, for each $t$, the model is defined as follows:
\begin{align}
   \hspace{-0.2cm} &\bigg(\dfrac{y_{t} - \hat{\mu}_{m_{t}}}{\hat{\sigma}_{m_{t}}}\bigg) = \phi_{1}^{(m_{t})}\bigg(\dfrac{y_{t-1} - \hat{\mu}_{m_{t-1}}}{\hat{\sigma}_{m_{t-1}}}\bigg) + \dots + \nonumber  \\ \hspace{-0.2cm}&\phi_{p_{m_{t}}}^{(m_{t})}\bigg(\dfrac{y_{t-p_{m_{t}}} - \hat{\mu}_{m_{t}-p_{m_{t}}}}{\hat{\sigma}_{m_{t}-p_{m_{t}}}}\bigg) + \psi^{(m_t)}\bigg(\dfrac{{\hat{A}}_{t} - \hat{\mu}^{(A)}_{m_t}}{\hat{\sigma}^{(A)}_{m_t}}\bigg) + \epsilon_{t}.  \label{PARpA_model}
\end{align}
where
\begin{align}
\hat{A}_{t} &= \frac{1}{S}\sum_{j = 1}^S y_{t-j} \\ 
\hat{\mu}_{m}^{(A)} &= \frac{1}{N_{m}}\sum_{j=1}^{N_{m}} \hat{A}_{jS + m} \\
\hat{\sigma}_{m}^{(A)} &= \sqrt{\frac{1}{N_{m}}\sum_{j=1}^{N_{m}} \big(\hat{A}_{jS + m} - \hat{\mu}_{m}^{(A)}\big)^2} \\
N_{m}~&\text{is the number of periods corresponding }  \nonumber\\
&\text{to month $m \in M$ in the historical data} \nonumber
\end{align}

In the official implementation overseen by the Brazilian energy authority, the autoregressive parameters for the PARp model are determined using the Yule-Walker equations \citep{GTMetodologiaCPAMP2022}. 
In the case of the PARp-A model, parameter estimation is similarly derived through an extension of the conventional Yule-Walker equations, tailored to accommodate the inclusion of an additional annual moving-average term
\citep{GTMetodologiaCPAMP2022}. Alternatively, these estimates can be obtained using Ordinary Least Squares (OLS), as done in \cite{PARp_PSR}.

After estimating the values for all model coefficients ($\phi's$ and $\psi's$) for each of the 12 normalized autoregressive models, the random errors in \eqref{PARp_model} or \eqref{PARpA_model} can be estimated by the residuals of the model. According to the official methodology, these errors are assumed to follow a three-parameter log-normal distribution, i.e., $\epsilon_{t}$  $\sim LN(\mu^{(\epsilon)}_{m_t}, \sigma^{(\epsilon)}_{m_t}, \lambda_{t})$. Because the errors have zero means, it is clear that the third parameter $\lambda_{t}$ should be negative to shift the standard positive (two-parameter) log-normal for this mean. Thus, while $\mu^{(\epsilon)}_{m}$ is set to zero and $\sigma^{(\epsilon)}_{m}$= $\hat{\sigma}^{(\epsilon)}_{m}$ for all $m\in M$, the procedure to obtain the third parameter $\lambda_{t}$ is nonstandard. Therefore, the third parameter is calculated to fulfill the following requirements for the final shifted distribution: 1) zero mean, 2) a given sample standard deviation, and 3) positivity for the time series. This third requirement produces a heuristic procedure detailed in the Appendix.

Finally, to estimate the spatial correlation among subsystems for generating probabilistic forecasts, the set of periodic correlation matrices, $\{\boldsymbol{\hat{U}}_{m}\}_{m\in M}$, is estimated based on the residuals of the models \citep{pereira1984stochastic}. %\footnote{In the Brazilian System Operator documentation, other heuristics are reported for calculating this correlation matrix, such as using the time series correlation for the residuals. However, due to the lack of explicit details, we selected the method described in the official methodology's technical notes and reference manuals.} 
Then, $\boldsymbol{\hat{B}}_{m}$ can be obtained as the Cholesky decomposition of $\boldsymbol{\hat{U}}_{m}$, enabling the simulation of correlated scenarios based on independent ones. For a detailed description of the $k$-step-ahead
(probabilistic and point) forecast procedure, we refer to the Appendix.

%The observed NIE data exhibit both highly asymmetrical and non-negative characteristics. To mitigate the risk of generating negative scenarios, the Brazilian official methodology, outlined in \cite{jardim2001stochastic}, employs a dynamically adjusted three-parameter log-normal distribution (see \cite{charbeneau1978comparison}). In this methodology, 

\section{Case study} \label{case_study}

In this section, we apply the metrics discussed in Section \ref{metrics} to evaluate the results obtained from the PARp, PARp-A, and benchmark models. The study used the official Brazilian dispatch software, NEWAVE (version 28.0.3) \citep{GTMetodologiaCPAMP2022}, to generate the NIE forecasts for both the PARp and PARp-A official models. The following steps were taken to obtain the data:

\begin{itemize}
\item The input data published by the Brazilian National System Operator for the NEWAVE software in October 2024 served as the basis for building a reference configuration for the system used throughout the study.
\item Expansion and any modifications to the hydroelectric plants were disregarded within the study horizon, ensuring the fixed base of NIE.
\item The study spanned from January 2011 to September 2024, with only the initial date of the study and the historical observed NIE provided by NEWAVE being altered in each iteration.
\item For each initial date, the NEWAVE software uses the official PARp-based forecast model to generate 2000 scenarios (chronologically linked paths), 60 months ahead, for the NIE time series for all Brazilian subsystems according to the methodology described in Section \ref{Parp:section} and in the Appendix.
\end{itemize}

It is noteworthy that the official implementation considers the complete dataset in the estimation process, from the beginning of the historical records (January 1931) until two years prior to the period in which the forecast is made. So, the official methodology disregards the last two years of observed NIE data. For instance, in January 2011, only historical data from January 1931 to December 2009 were utilized. However, in this work, we considered all observed data in the estimation step. This approach ensures that the bias assessment is isolated from the influence of disregarding the last two years, thus incorporating the most recent information available.

%For the PARp and PARp-A models, $|\Omega|$ was set to 2000. 
We explore the flexibility of using different values for $n_k$ (the number of out-of-sample k-steps-ahead forecasts) for different values of $k$. For example, the forecast made with data up to December 2010 for January 2011 is projected 24 steps ahead. Conversely, the forecast made with data up to August 2024 for September 2024 is only 1 step ahead, owing to the consideration of observed data up to September 2024. In this context, we have $n_1 = 152$, $n_2 = 151$, so on, until $n_{24} = 129$. All input and output data are available in \cite{CaseStudyData}.

\subsection{PARp and PARp-A Model Forecasts}

\begin{figure}[!ht]
\includegraphics[scale = 0.37, trim={6.2cm 2.3cm 4.9cm 1.6cm},clip]{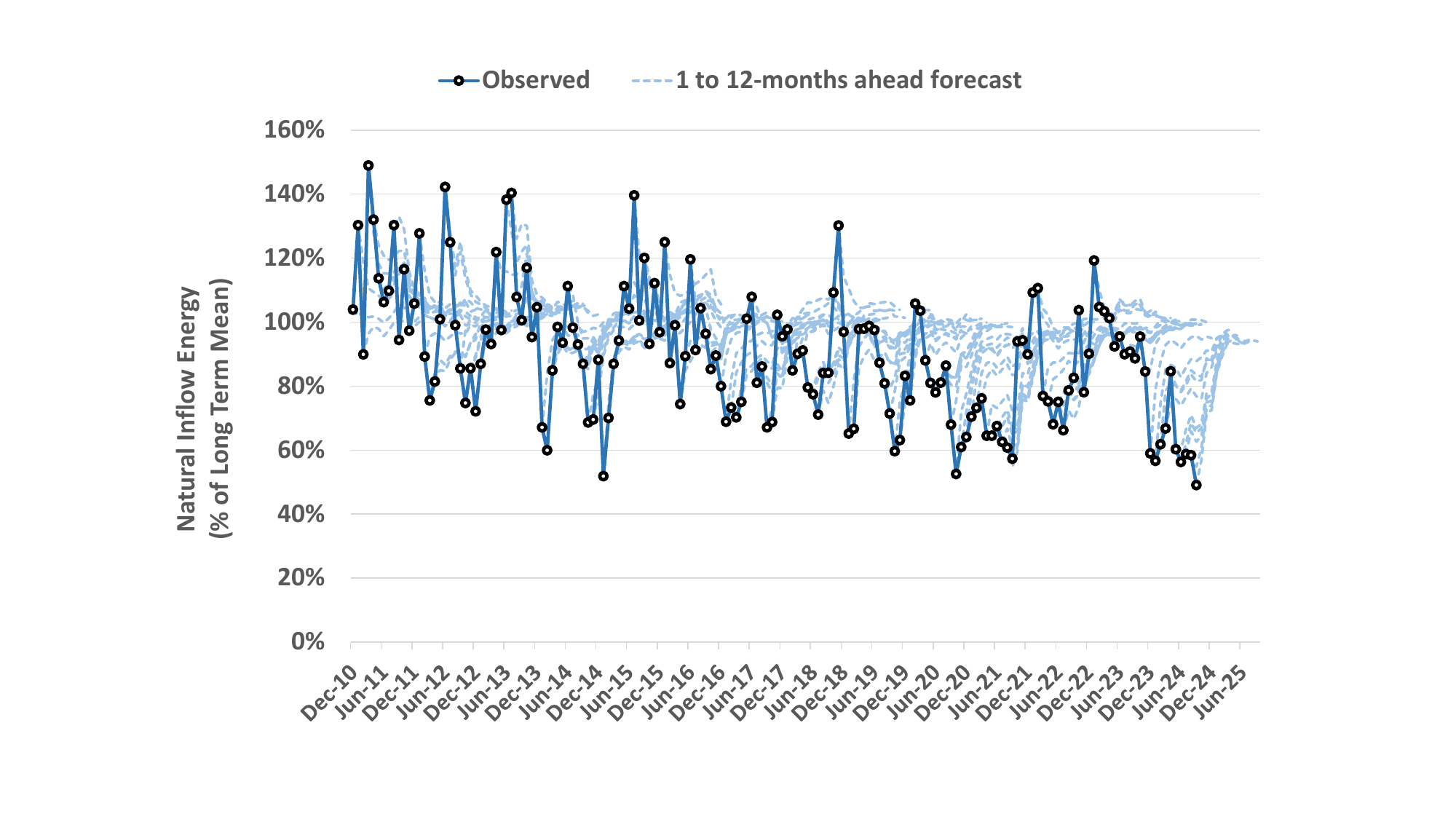}
\centering
\caption{PARp Forecasts and Observations of NIE for the SE subsystem in \% of the seasonal long-term mean.}
\label{PARp_SE}
\end{figure}

\begin{figure}[!ht]
\includegraphics[scale = 0.37, trim={6.1cm 2.3cm 4.9cm 1.6cm},clip]{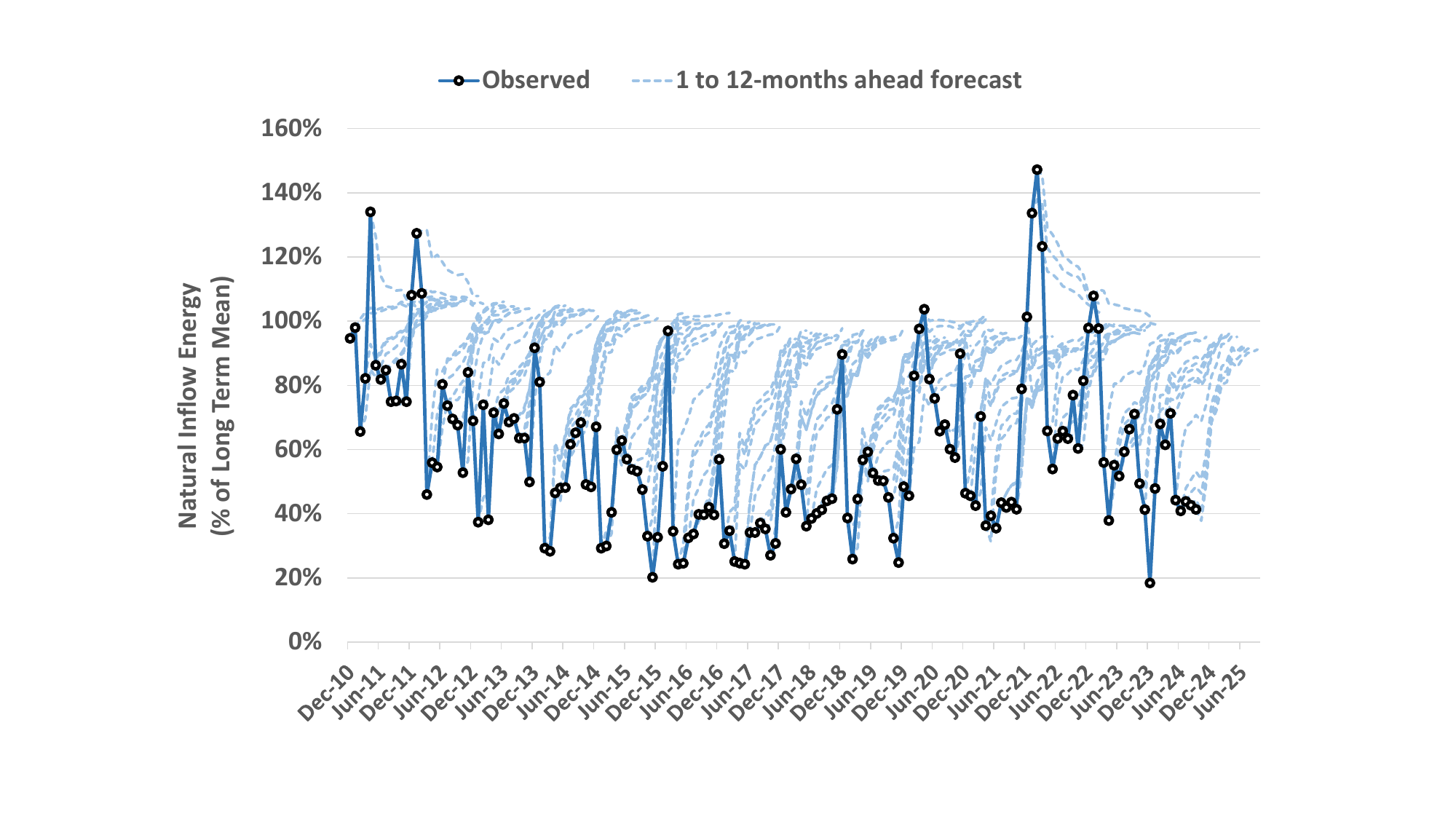}
\centering
\caption{PARp Forecasts and Observations of NIE for the NE subsystem in \% of the seasonal long-term mean.}
\label{PARp_NE}
\end{figure}

Figures \ref{PARp_SE} and \ref{PARp_NE} depict the forecasts against observed values in percentage of the seasonal LTM (calculated for each month using the complete historical dataset). In these figures, the forecasts are obtained utilizing the PARp model for the SE (\ref{PARp_SE}) and NE (\ref{PARp_NE}) subsystems. The continuous blue line with circular markers represents the observed values, while each lighter dashed line depicts the first 12-step-ahead point forecasts made from each period in the evaluation horizon.\footnote{It is important to note that each point forecast within the rolling-window evaluation horizon (2011–2024), as depicted in Figures 1 to 4, was generated by the official software considering the PARp or PARp-A models, both estimated with the entire historical dataset available up to the time the forecast was made, i.e., from January 1931 up to a given period $t\in\{Jan/2011, ..., Sep/2024\}$. Moreover, the 12-step-ahead point forecasts were derived by averaging 2000 synthetically generated scenarios, each of which composed of 12 chronologically simulated paths as per the Appendix. Thus, each forecast curve represents the first moment of the 12 conditional predictive distributions obtained for each period in the evaluation horizon over the subsequent 12 months.} It can be observed that both subsystems quickly revert to the long-term mean, as expected (as the PARp family of models are all mean-reverting models). Despite the well-known mean-reverting property inherent to the official forecast models, a notable negative trend is evident in the analyzed NIE time series (Figures \ref{PARp_SE} and \ref{PARp_NE}), with forecasts predominantly exceeding actual observed values throughout the analyzed period, resulting in a clear positive bias (forecasts higher than observations on average). These facts offer insights into why the planning policies derived from the official methodology exhibited higher hydro dispatches than anticipated, as discussed in Section \ref{Introduction}.


\begin{figure}[!ht]
\includegraphics[scale = 0.37, trim={6.2cm 2.3cm 4.8cm 1.6cm},clip]{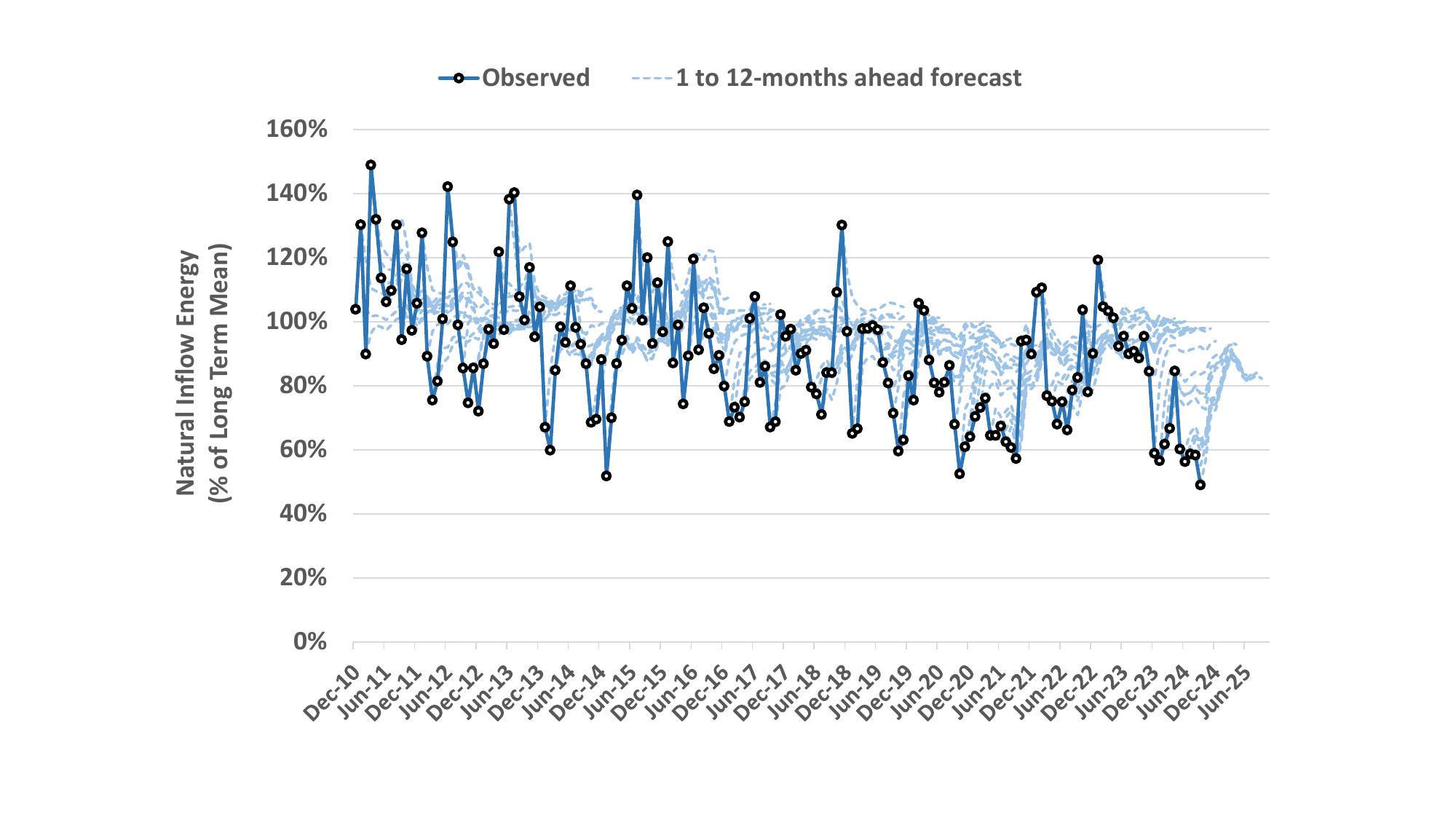}
\centering
\caption{PARp-A Forecasts and Observations of NIE for the SE subsystem in \% of the seasonal long-term mean.} 
\label{PARpA_SE}
\end{figure}

\begin{figure}[!ht]
\includegraphics[scale = 0.37, trim={6.2cm 2.3cm 4.7cm 1.6cm},clip]{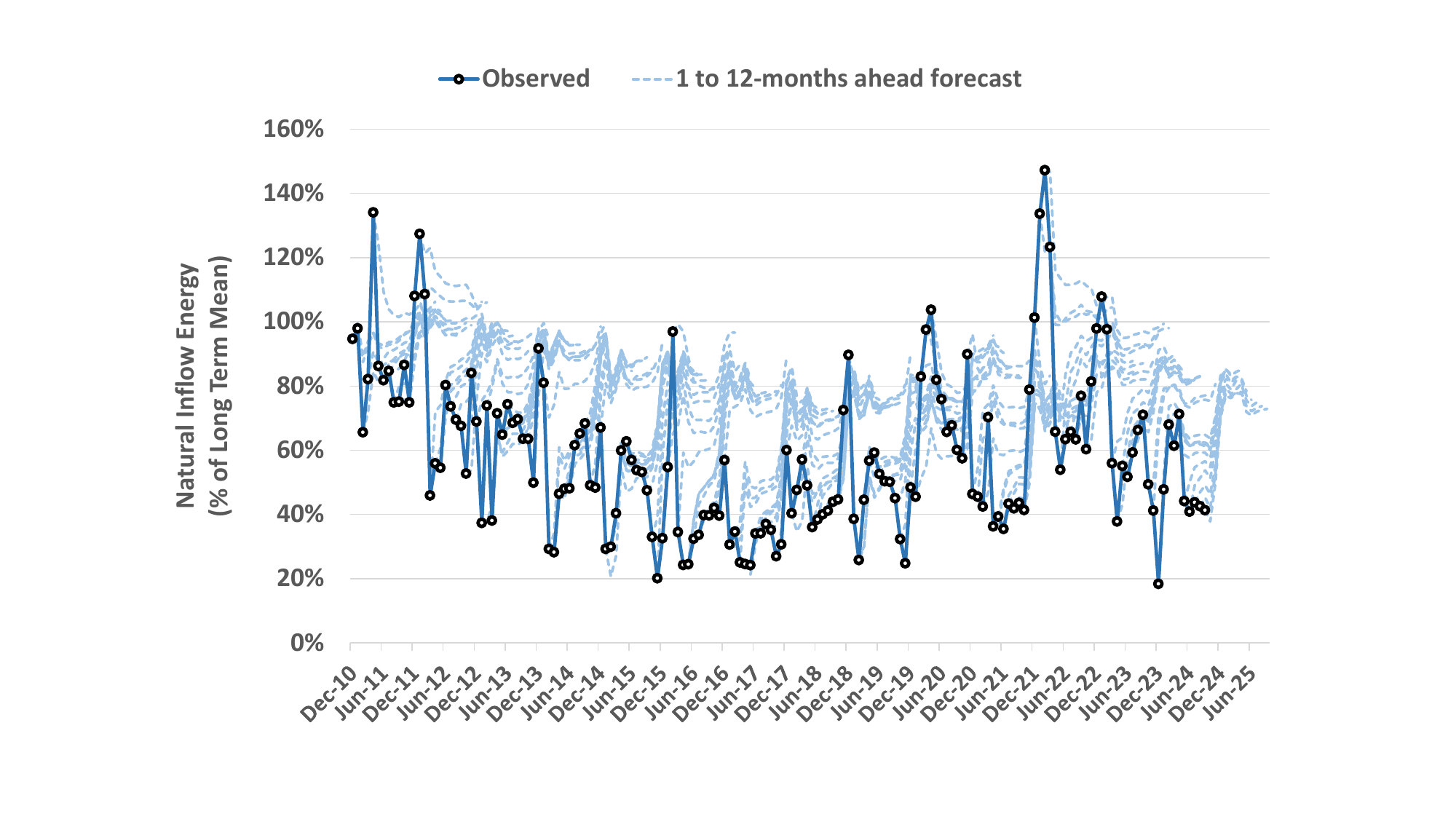}
\centering
\caption{PARp-A Forecasts and Observations of NIE for the NE subsystem in \% of the seasonal long-term mean.}
\label{PARpA_NE}
\end{figure}

Figures \ref{PARpA_SE} and \ref{PARpA_NE}, utilizing forecasts from the PARp-A model, exhibit a similar pattern. In this instance, regression to the long-term mean occurs more gradually, yet a prevailing trend persists, with forecasts often surpassing the observed values.


\subsection{PARp and PARp-A forecast bias}

\begin{figure}[!ht]
    \includegraphics[scale = 0.36, trim={5.3cm 1.8cm 3.6cm 0.5cm},clip]{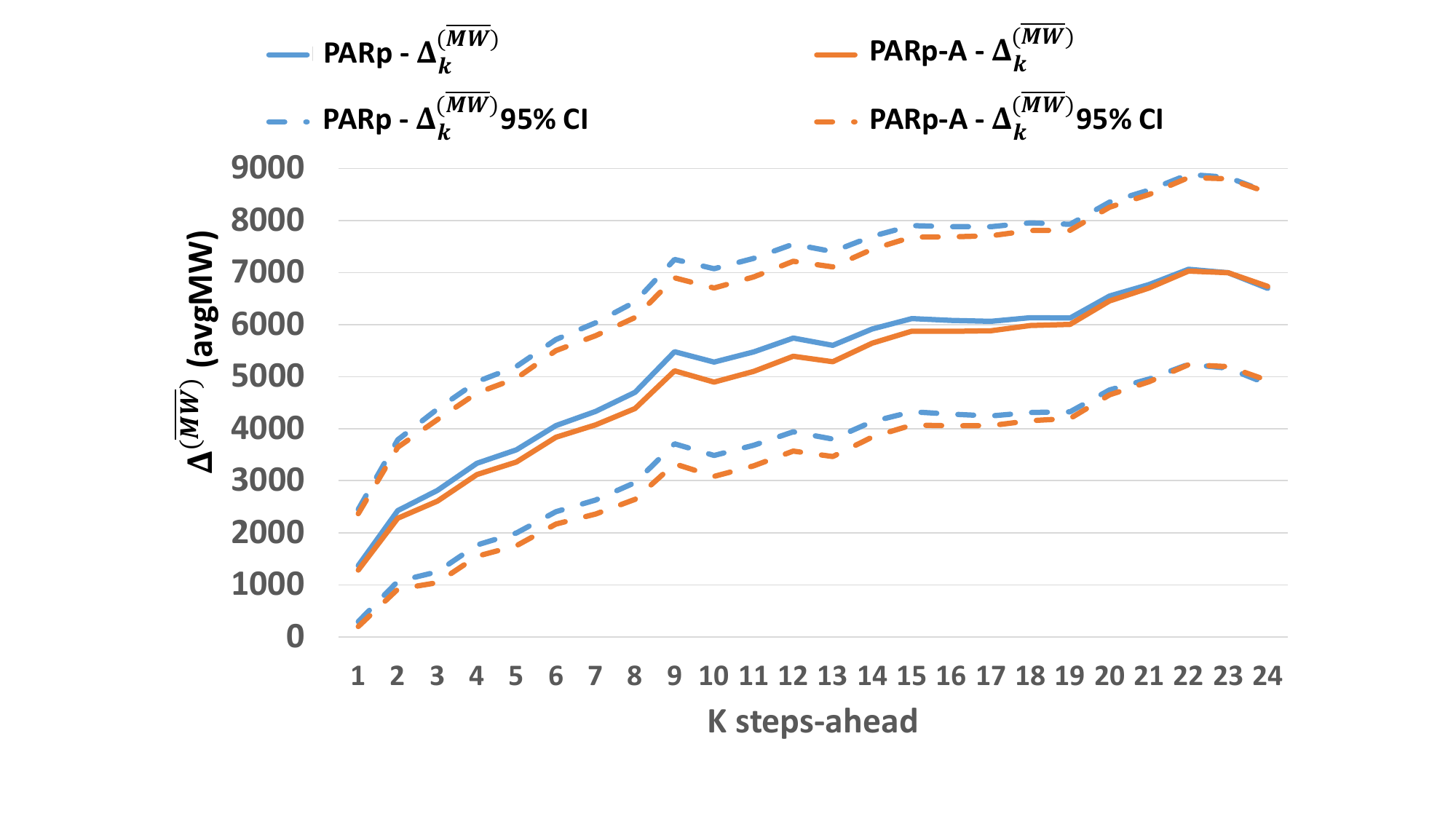}
\centering
\caption{Southeast NIE Forecasts bias in average MW and 95\%--confidence interval (2.5 and 97.5\% quantiles) for PARp and PARp-A}
\label{vies_PARp_PARpA_SE}
\end{figure}

\begin{figure}[!ht]
    \includegraphics[scale = 0.36, trim={5.3cm 1.8cm 3.6cm 0.5cm},clip]{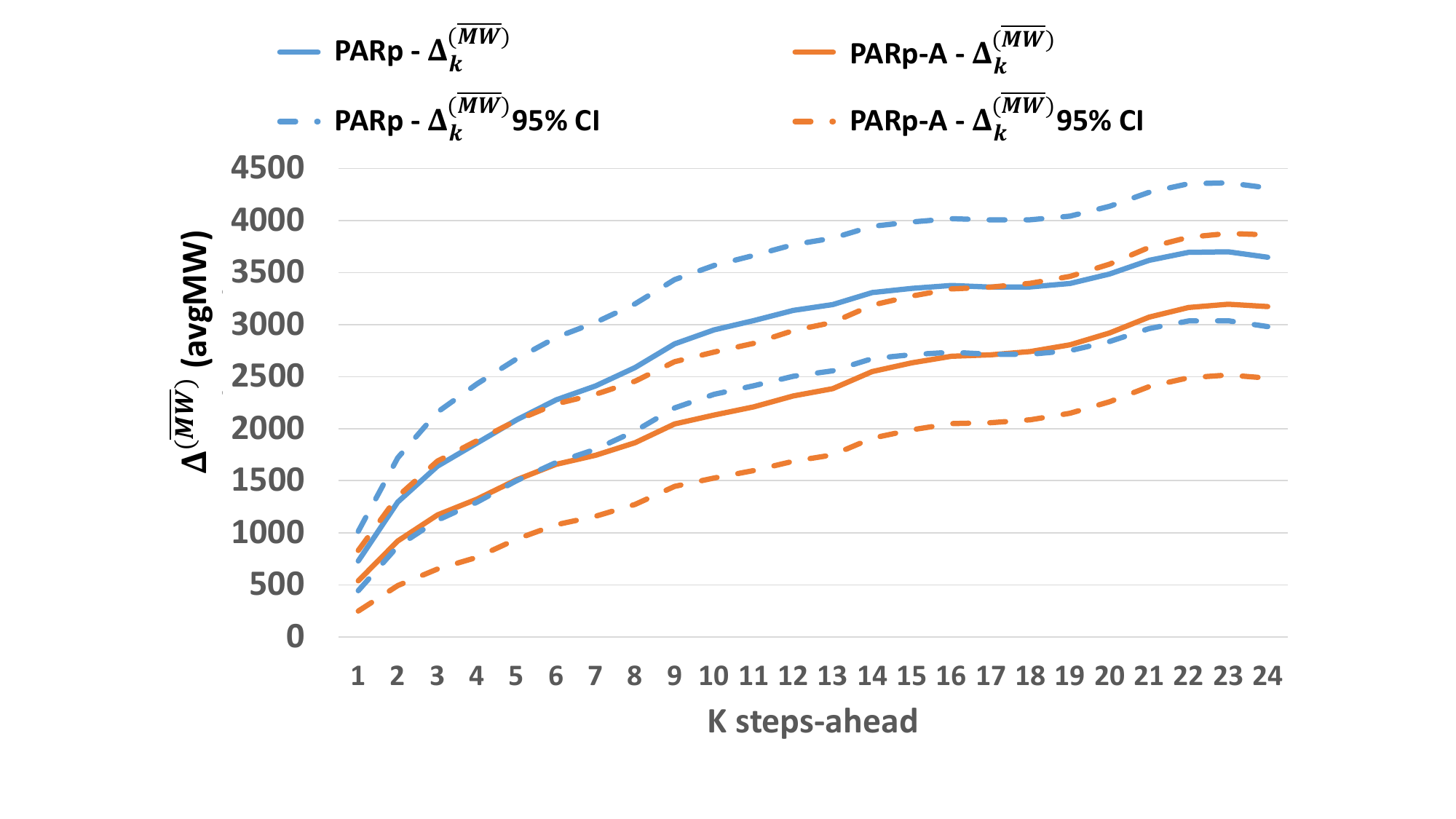}
\centering
\caption{Northeast NIE Forecasts bias in average MW and 95\%--confidence interval (2.5 and 97.5\% quantiles) for PARp and PARp-A}
\label{vies_PARp_PARpA_NE}
\end{figure}

Figures \ref{vies_PARp_PARpA_SE} and \ref{vies_PARp_PARpA_NE} depict the evaluated bias for the PARp and PARp-A models in the SE and NE subsystems according to equation \eqref{bias_MWm}. It is clear that the bias is consistently positive over time for both models, indicating that, on average, the forecasts exceeded the observations. \emph{This observation provides compelling evidence that these two stationary models, devoid of additional structure or external explanatory signals (as studied in \cite{pina2017optimizing}), fail to capture the downward pattern empirically observed in the analyzed time series.}
Furthermore, it is clear that forecast errors escalate with longer forecast horizons. While there is nearly no distinction between the results obtained from the PARp and PARp-A models for the SE subsystem, a reduction in bias is discernible for the NE subsystem, although the bias remains positive.

\begin{figure}[!ht]
    \includegraphics[scale = 0.36, trim={5.3cm 1.8cm 3.6cm 0.5cm},clip]{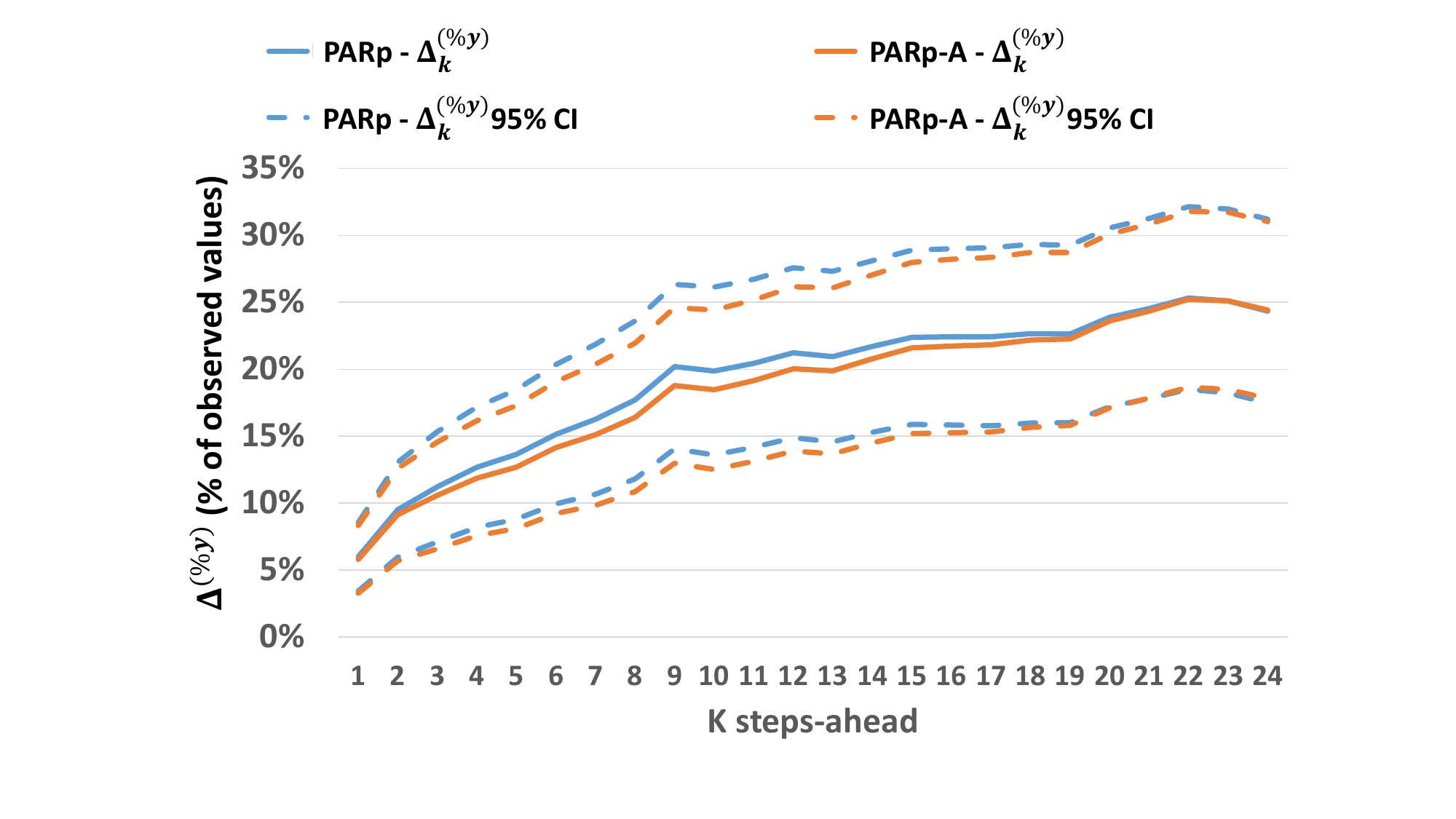}
\centering
\caption{Southeast NIE Forecasts bias in percentage of observed values and 95\%--confidence interval (2.5 and 97.5\% quantiles) for PARp and PARp-A}
\label{vies_PARp_PARpA_SE_obs}
\end{figure}

\begin{figure}[!ht]
    \includegraphics[scale = 0.36, trim={5.2cm 1.8cm 3.6cm 0.5cm},clip]{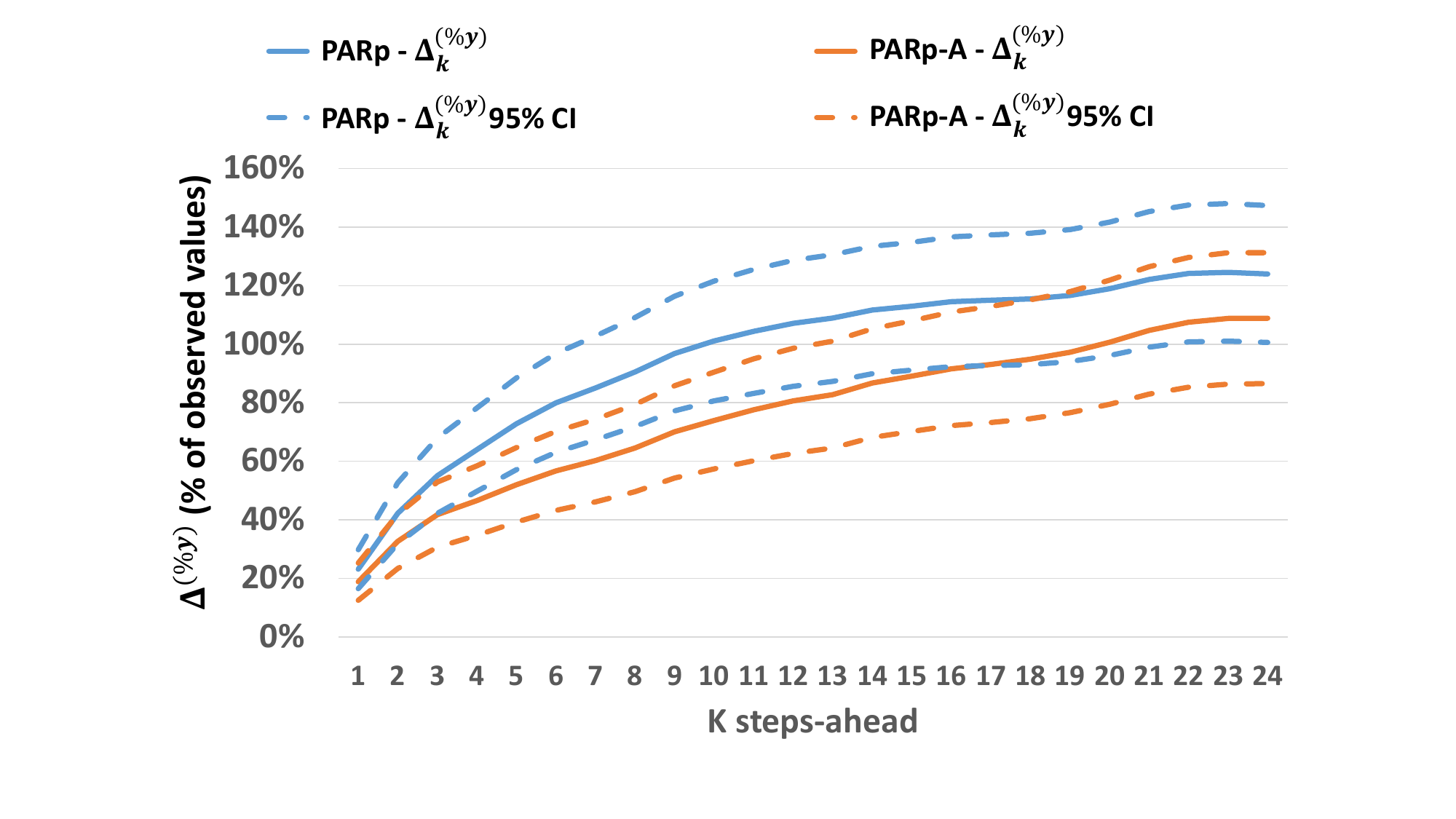}
\centering
\caption{Northeast NIE Forecasts bias in percentage of observed values and 95\%--confidence interval (2.5 and 97.5\% quantiles) for PARp and PARp-A}
\label{vies_PARp_PARpA_NE_obs}
\end{figure}

Figures \ref{vies_PARp_PARpA_SE_obs} and \ref{vies_PARp_PARpA_NE_obs} 
compare the evaluated bias and their (95\%) confidence intervals for the PARp and PARp-A models in the SE and NE subsystems according to equation \eqref{bias_y}. It is notable that for forecasts 12 steps ahead in the SE subsystem, the bias exceeds 15\% of the observed values, and it surpasses 20\% of the observed values for forecasts 24 steps ahead. The situation is even more pronounced in the NE subsystem, with the bias for forecasts 24 steps ahead exceeding 100\%. % of the observed values for both the PARp and PARp-A models.

Based on the results presented in this section, the following important conclusion remarks can be drawn: 
\begin{enumerate}
    \item The official forecast procedure, which uses as input the complete NIE historical dataset to estimate the PARp-A model and generates as output multi-steps-ahead forecasts, presents statistically significant bias for both the SE and NE subsystems.
    \item Specifically, bias levels were found to be 1.28, 3.83, 5.39, and 6.73 average GW for 1-, 6-, 12-, and 24-step-ahead forecasts in the Southeast subsystem, and 0.54, 1.66, 2.32, and 3.17 average GW in the Northeast subsystem. This indicates a consistent optimistic bias in the official NIE forecasts for the two main subsystems of the Brazilian power system over the last 12 years. Furthermore, a pattern of increasing bias was identified with the number of forecast steps ahead.
    \item The PARp-A model slightly reduces the bias compared to the PARp model. This reduction is more relevant for the NE subsystem.
    \item The mean-reverting property of stationary PARp-A models and the increasing bias with the number of steps ahead with which the forecast is made provide relevant evidence on the lack of structure in the official model to capture the trends observed in historical data. 
\end{enumerate}

\subsection{Benchmarking the official PARp-A forecast bias}

%\section{Benchmark models} \label{benchmarks}

To provide alternative references for assessing bias in the official forecasting procedure, we benchmark the estimated bias against variants of the PARp-A model and {with other typical classes of models. The PARp-A variants are used to benchmark the official methodology against similar models that adhere to the same assumptions required by the SDDP application, namely, models that are linear in their parameters and capable of generating synthetic scenarios with strictly positive values. Three PARp-A variants are introduced, each employing a different form of bias correction. The other models, although not completely compliant with the SDDP requirements, are included as reference benchmarks. \emph{However, \textbf{It is important to emphasize} that the purpose of this benchmarking is not to identify an alternative unbiased model---this remains an open research question that requires further investigation using additional metrics and analyses. Rather, the goal is to provide insights and references regarding the bias identified in the previous section, the main focus of this paper.}} \\ 

%\noindent \textbf{Recent monthly median} --- This benchmark model is a stabilized version of the Seasonal Naive model. So, when forecasting period $t+k$, we use the median of the last five observed values for the same month of $t+k$. It also aims to account for recent information and forget long-term profiles while using a few years to stabilize the forecast based on the median value.

\noindent \textbf{PARp-A variant with shorter memory (estimation process using only the last $J$ years)} --- The official implementation of the PARp-A model in Brazil utilizes all observed data since January 1931 in the estimation step. However, this benchmark provides a reference within the same framework of the official methodology. By excluding data older than $J$ years before the last value observed in the estimation process in each window of the rolling horizon test, this benchmark aims to exclude bias due to old long-term patterns. This benchmark is available in the PARp implementation shared at \cite{PARp_PSR}. \\

\noindent \textbf{PARp-A variant with short-term focus (estimation process emphasizing the most recent year with weight $w$)} --- This benchmark explores the option of emphasizing the most recent year by multiplying the error term of the most recent 12 months by a weight $w > 1$ in the estimation process. In this case, we still utilize all observed data since January 1931, as is done in the official implementation in Brazil, but emphasize the fitting to the most recent data. The implementation of this benchmark model is also available in \cite{PARp_PSR}.\\

\noindent \textbf{PARp-A variant with Adjusted Long-Term Mean (ALTM)} --- In this variant, we introduce a different bias‐reduction scheme for the PARp‐A estimation. The main idea is to adjust the local LTM of the log-transformed historical data to the most recent local LTM estimates. The steps are summarized as follows:

\begin{enumerate}
  \item Apply the log transformation to the available observed data: $z_t = \ln(y_t)$.
  \item Evaluate the moving average time series with 12 months (was the best result tested): $A_t^{(z)} = \frac{1}{M} \sum_{i=0}^{M-1} z_{t-i}$.
  \item Set the target LTM: $A^{(z)}_{\mathrm{target}} = A^{(z)}_T$.
  \item Normalize the log historical data to the most recent local estimates of the LTM: $\tilde z_t = z_t - A^{(z)}_t + A^{(z)}_{\mathrm{target}}$.
  \item Apply the inverse transformation (exponential) to bring the adjusted data to the original level: $\tilde y_t = \exp\bigl(\tilde z_t\bigr)$.
  \item Estimate the PARp-A model as usual for the complete adjusted historical data $\tilde y_t$.
\end{enumerate}

As described in the Appendix, all PARp-A benchmarks, as well as the official PARp-A model, use a log-normal distribution to generate scenarios and forecasts. Thus, we tested the following benchmarks with and without applying a log transformation to the data. Since all models performed better on the transformed data, we applied this transformation to all of them. Nevertheless, all reported results and performance metrics are presented on the original (NIE) scale.\\

\noindent \textbf{Classic SARIMA model} --- This benchmark model provides us with the classical benchmark. We utilize the SARIMA model as implemented in \cite{SARIMA}. Fits are made using the \textit{auto.arima} function\\

\noindent \textbf{XGBoost} ---  
We choose a gradient-boosting regression model based on XGBoost \citep{XGBoost} to explore how these techniques handle nonlinear patterns in time series. We construct an explanatory feature vector including observations from the past 24 months and the moving average over the past 60 months to provide both short- and long-term context. We set the maximum tree depth to 8 and a learning rate of 0.3\\

\noindent \textbf{Chronos} ---  
The Chronos model \citep{ansari2024chronos} is based on Transformer architectures optimized for time-series forecasting, leveraging attention mechanisms to capture long-range dependencies. We use the Chronos T5 (Base) variant with 200 samples. \\

\noindent \textbf{Prophet} ---  
Prophet \citep{taylor2018forecasting} combines an additive model of trend and seasonality. We employ the default multiple-seasonality configuration and 95\% confidence intervals. \\

\noindent \textbf{Seasonal naïve} --- This is a nonparametric model that focuses on providing the most naive forecast possible with recent information. When forecasting period $t+k$ in the rolling-window scheme, this model simply considers the forecast to be equal to the last observed value of the same month, i.e., $y_{t+k-12}$. Although extremely naive and with a high error rate, it provides us with the most recent seasonal information immunized against the need to model complex long-term effects. As will be seen, this benchmark model plays a relevant role as a reference in bias monitoring under the presence of long-term trends or cycles. \\
%\noindent \textbf{Score driven model} --- We consider a score-driven model (p,q) with a log-normal distribution, where the mean parameter varies over time. This benchmark offers a reference based on a non-gaussian model where the lognormal has its parameter varying over time. We use a scaling factor of $1$ and incorporate lags of $1$, $2$, $11$, and $12$. The forecasting is performed using the implementation in \citep{bodin2020scoredrivenmodels}. For the estimation step, we also use the most recent 30 years in the estimation process for each step $t$ of the rolling-window scheme. Note that this benchmark keeps the same number of data points considered in the SARIMA and PARp-A($J=30$) in the estimation process for comparison purposes.  

%\subsection{Structural time series model}

%In this setup, the time series is decomposed into both stochastic trend and seasonal components, estimated through Kalman filtering. For this, we used the implementation presented in \citep{SaavedraBodinSouto2019}. The forecast by the structural model is advantageous because the model adapts to the most recent realizations, and stationarity is not required. In this case, we also use the logarithm of the series and consider the most recent 30 years in the estimation process for each step $t$ of the rolling-window scheme.

\begin{table}[!ht]
\caption{Mean bias comparison for PARp-A against benchmark models for the Southeast subsystem (avgGW)}
\centering
\resizebox{\columnwidth}{!}{%
\begin{tabular}{ccccccc}
\textbf{}                                        & \textbf{K = 1} & \textbf{K = 6} & \textbf{K = 12} & \textbf{K = 24} & \multicolumn{2}{c}{\textbf{cumulative bias}} \\ \hline
\textbf{Official PARp-A}                         & 1.28           & 3.83           & 5.39            & 6.73            & 110.07           & \textbf{\%PARp-A}          \\ \hline
\textbf{Seasonal Naïve}                          & 1.16           & 1.45           & 1.65            & 1.77            & 35.21            & 32\%                       \\ \hline
\textbf{PARp-A ($J = 30$)}                       & 2.73           & 6.64           & 6.69            & 7.22            & 143.58           & 130\%                      \\
\textbf{PARp-A ($J = 50$)}                       & 1.37           & 5.29           & 6.15            & 7.44            & 128.00           & 116\%                      \\
\textbf{PARp-A ($J = 70$)}                       & 0.95           & 4.51           & 4.84            & 6.05            & 103.30           & 94\%                       \\ \hline
\textbf{PARp-A ($w = 2$)}                        & 1.10           & 4.73           & 5.00            & 6.08            & 106.92           & 97\%                       \\
\textbf{PARp-A ($w = 4$)}                        & 1.10           & 4.68           & 4.93            & 6.04            & 105.85           & 96\%                       \\
\textbf{PARp-A ($w = 11$)}                       & 0.93           & 4.23           & 4.56            & 5.81            & 98.70            & 90\%                       \\ \hline
\multicolumn{1}{l}{\textbf{ALTM PARp-A}} & 0.66           & 2.02           & 1.73            & 2.35            & 38.18            & 35\%                       \\ \hline
\textbf{SARIMA}                                  & 1.64           & 3.22           & 3.21            & 3.63            & 71.04            & 65\%                       \\ \hline
\textbf{XGBoost}                                 & 0.80           & 2.11           & 3.28            & 1.87            & 51.83            & 47\%                       \\ \hline
\textbf{Prophet}                                 & 2.31           & 3.02           & 3.43            & 4.21            & 75.85            & 69\%                       \\ \hline
\textbf{Chronos}                                 & 1.04           & 4.13           & 4.25            & 3.85            & 89.50            & 81\%                      
\end{tabular}%
}
\label{bias_benchmarks_SE}
\end{table}

\begin{table}[!ht]
\caption{Mean bias comparison for PARp-A against benchmark models for the Northeast subsystem (avgGW)}
\centering
\resizebox{\columnwidth}{!}{%
\begin{tabular}{ccccccc}
\textbf{}                    & \textbf{K = 1} & \textbf{K = 6} & \textbf{K = 12} & \textbf{K = 24} & \multicolumn{2}{c}{\textbf{cumulative bias}} \\ \hline
\textbf{Official PARp-A}     & 0.54           & 1.66           & 2.32            & 3.17            & 49.03            & \textbf{\%PARp-A}          \\ \hline
\textbf{Seasonal Naïve}      & 0.09           & 0.21           & 0.23            & 0.30            & 5.27             & 11\%                       \\ \hline
\textbf{PARp-A ($J = 30$)}   & 0.51           & 1.25           & 1.52            & 2.03            & 33.84            & 69\%                       \\
\textbf{PARp-A ($J = 50$)}   & 0.30           & 1.27           & 1.57            & 2.29            & 34.54            & 70\%                       \\
\textbf{PARp-A ($J = 70$)}   & 0.49           & 1.70           & 2.00            & 2.80            & 44.73            & 91\%                       \\ \hline
\textbf{PARp-A ($w = 2$)}    & 0.47           & 1.80           & 2.15            & 2.98            & 47.64            & 97\%                       \\
\textbf{PARp-A ($w = 4$)}    & 0.46           & 1.75           & 2.06            & 2.89            & 46.11            & 94\%                       \\
\textbf{PARp-A ($w = 11$)}   & 0.39           & 1.50           & 1.74            & 2.58            & 40.19            & 82\%                       \\ \hline
\textbf{ALTM PARp-A} & - 0.13         & 0.02           & 0.11            & - 0.03          & - 0.02           & 0\%                        \\ \hline
\textbf{SARIMA}              & 0.37           & 0.60           & 0.67            & 0.72            & 14.55            & 30\%                       \\ \hline
\textbf{XGBoost}             & - 0.09         & 0.20           & 0.16            & 0.30            & 3.83             & 8\%                        \\ \hline
\textbf{Prophet}             & - 0.20         & - 0.11         & - 0.07          & 0.08            & - 0.98           & -2\%                       \\ \hline
\textbf{Chronos}             & - 0.15         & - 0.05         & - 0.06          & 0.01            & - 1.99           & -4\%                      
\end{tabular}%
}
\label{bias_benchmarks_NE}
\end{table}

Tables \ref{bias_benchmarks_SE} and \ref{bias_benchmarks_NE} showcase the forecast bias {in average gigawatts} evaluated for 1, 6, 12, and 24 months ahead for the SE and NE subsystems, respectively. Additionally, these tables include the bias for the 24-month ahead cumulative forecast. {Note that the values presented in the first row of Tables \ref{bias_benchmarks_SE} and \ref{bias_benchmarks_NE}, which refer to the bias of the Official PARp-A model, correspond to those shown by the continuous orange curves in Figures \ref{vies_PARp_PARpA_SE} and \ref{vies_PARp_PARpA_NE}, scaled by a factor of 1,000.} From these tables, we can see that the biases previously depicted {(in average megawatts) for the Official PARp-A model in Figures \ref{vies_PARp_PARpA_SE} and \ref{vies_PARp_PARpA_NE} are also present in most of the benchmarks with different sizes. }

{It is relevant to clarify that the cumulative inflow metric --- representing the total volume of water (or energy) forecasted over the next two years\footnote{{For each step $t$ in the rolling window test, the 24-month ahead cumulative forecast is calculated by summing up the inflow forecasted values from 1 to 24 steps, i.e., $\hat{y}_{[24]|y_{[t]}} = \sum_{k=1}^{24} \hat{y}_{t+k|y_{[t]}}$. Then, the cumulative forecast is contrasted with the observed cumulative inflow in the same horizon, resulting in a cumulative error, $\hat{\varepsilon}_{t,[24]} = \sum_{k=1}^{24}\hat{\varepsilon}_{t,k}$, that is used to assess the cumulative bias.}} --- is frequently used by practitioners due to the multi-year nature of the opportunity cost associated with storing water for future stages in hydrothermal operational planning. The rationale in this metric is that although a monthly forecast might deviate significantly from observed values, analyzing cumulative inflows provides relevant insights into whether the forecasting model is consistently biased in estimating total water/energy inflows. Storage reservoirs could theoretically mitigate alternating monthly forecast errors. However, results in Tables \ref{bias_benchmarks_SE} and \ref{bias_benchmarks_NE} reveal a persistent bias that does not average out over time.}

Additionally, it is evident that reducing the size of the historical data window does not yield improvement in results compared to PARp-A in the SE subsystem, but it demonstrates significant improvement for the NE subsystem. Similarly, emphasizing the last year of historical data in the estimation process shows similar patterns. The naive benchmark, except for $K=1$ in the Southeast, exhibits the lowest bias among all models. {
In both subsystems, the ALTM PARp-A model achieves the lowest cumulative bias by adjusting the historical data to the most recent log-local LTM, confirming that focusing on recent years significantly improves forecast performance. Likewise, the machine-learning models uniformly reduce bias compared to the baseline PARp-A, with Prophet standing out alongside ALTM for overall accuracy.}

Analysis of the bias of the benchmark models reveals a discernible pattern: the higher the relevance of the recent data in the estimation process of stationary models, the lower the bias. This pattern is evident when reducing the size of historical data used in the estimation process, increasing the number of times the last data is emphasized in the error minimization, and reaching its lowest bias in the extreme case of the seasonal naïve model, where only the last data corresponding to the same target month is utilized for forecasting all the 24 steps ahead. 

{Additionally, among the PARp-A variants, the ALTM PARp-A, which renormalizes the whole historical data to the local mean, exhibited the lowest bias within the testing horizon. It was also the best-performing model among all benchmarks in the Northeast and the second-best in the Southeast --- behind only the seasonal naïve model, which is not probabilistic. Finally, XGBoosting and Prophet performed consistently well for both subsystems.} These findings highlight a lack of structure in the official model to capture the negative trend observed in the last decade, which is better captured in the benchmark models through non-parametric adjustments in the relevance of recent data. %This fact corroborates the third conclusion remark drawn in the previous Section. 

\section{Discussion and Recommendations}

%The reliance of hydrothermal power systems on inflow forecasts is major. Therefore, biased forecasts, i.e., upward-shifted predictive distributions, have the potential to contaminate all subsequent planning activities as well as the price formation processes with optimistic bias. 

It is important to emphasize that this study does not aim to provide a definitive diagnosis of the official models. However, the results emphasize the necessity for further investigation and call for a new governance and monitoring process by local authorities responsible for official inflow forecasts in Brazil. It may also provide relevant insights for other hydro-dependent countries relying on similar forecast models and dispatch processes. 

It is relevant to highlight that the PARp-A model was selected as the official model due to its linear nature, which aligns with the hypotheses of SDDP, allowing it to generate inflow scenarios based on predicted distributions and derive the optimal operating policy. However, the rolling-window study reveals that PARp-A first-moment dynamics are incompatible with the apparent non-stationarity of the NIE data observed in the two main Brazilian reservoirs over the last decade. This is not only corroborated by Figures 1 to 4, but also by the statistically significant levels of optimistic bias quantified in Tables I and II. Nevertheless, although the causes of these effects are not investigated in this paper, there is a relevant piece of literature devoted to studying these effects.

%----------------------------------------
%ABOUT NEW TRENDS------------------------
%----------------------------------------
For instance, droughts in Brazil have been a recurrent topic in specialized literature in recent years. In recent years, most regions of Brazil experienced moderate to extreme drought, marking the most severe period observed for most areas \citep{geirinhas2021recent, marengo2022drought, de2024longest}. Furthermore, \cite{cuartas2022recent} indicates significant reductions in precipitation in relevant river basins associated with the Brazilian power system supply.

Regarding the causes, they have yet to be investigated and may encompass a range of factors, from irrigation to climate, among others. Interestingly, despite the well-known relationship between the rainfalls over the Brazilian Northeast and the position of the Intertropical Convergence Zone (see \cite{ mao2022phase}), recent studies have demonstrated that an indirect coupling effect between ENSO events and North Atlantic sea surface temperatures anomalies ---marked by an increased phase coherence--- has become more prominent in recent decades, reinforcing drought conditions over the Brazilian Northeast \citep{mao2022phase}. This coupling effect appears to have intensified in a warming climate, as strong ENSO events increasingly modulate Atlantic sea surface temperature variability. In addition, climate projections suggest that the Brazilian Northeast is likely to be one of the most affected regions by climate change and experience substantial hydroclimatic stress, with significant reductions in precipitation, soil moisture, and runoff by the end of the century \citep{cook2020twenty}. 

This scenario, although not fully conclusive, provides relevant evidence that the reported decade-long reduction in inflow profiles conflicts with the long-term mean-reverting property of the official forecasting model (PARp-A). This partially explains why this model may be exhibiting an optimistic bias --intuitively, reverting to the long-term mean from continuously low inflow states results in an optimistic (positive) bias.
%----------------------------------------

Given that inflows constitute the primary generation resource of the Brazilian power system, the generation of optimistically biased inflow scenarios, regardless of their underlying causes, can reduce water values, consequently leading to an optimistic and risky utilization of water resources. Therefore, we argue that, despite the uncertainty about which external drivers affect inflow time series and how, the compatibility between the model hypotheses and the data should be monitored to avoid biased results and other possible distortions. Within this context, given the complexity of the inflow time-series patterns and the high dependence of dispatch orders on the inflow forecasts, a new governance for both data and models is needed.

In this regard, \emph{we recommend that local authorities implement new monitoring procedures for the forecasting models used in assessing water opportunity costs (water values). Monthly monitoring of biases in 1- to 24-step-ahead forecasts represents a crucial first step, which should be complemented by additional error diagnostics and benchmark models to ensure forecasting accuracy. 
A second and crucial step is to monitor the impact of the reported inflow forecast bias on both short- and long-term operation.}

\emph{The isolated assessment of the impact of optimistic forecast bias on the actual operation of the Brazilian power system is indeed a complex undertaking and is part of the authors’ ongoing research efforts. {However, monitoring biases in forecasted operational outcomes (given by the planning models) against actual real-time operations is not a complex action. Such monitoring is essential to ensure timely actions that preserve power system reliability in the face of emerging climate patterns and other unforeseen effects. This can be readily implemented to help detect and alert authorities about systematic biases affecting the water values assessment. In this regard, a critical challenge this study faced in expanding on the operational impacts was the lack of an open-source dispatch ``mirror model". An open-source twin model that incorporates nearly all the features required to dispatch the system is a best practice for providing transparency, also aligning with current digitalization trends (see \cite{digital} and \cite{irena}). In addition, it unlocks the potential of the academic community to help improve current industry practices and the state-of-the-art (see some discussion in chapter 6.5 in \cite{Velloso})}.}

On the forecasting side, the current procedure used in the official models to generate non-negative forecasts introduces a nonlinear dependency in the third parameter of the log-normal distribution. Further research is needed to determine whether this issue compromises the convergence of the SDDP method or the accuracy of the probabilistic forecast. The nonlinear dependency used in the official SDDP implementations violates the SDDP hypotheses (see \cite{MATOS20121443}). Therefore, the purported linearity used to justify the PARp family of models should be revised, as the official simulation (forecast) procedure creates a nonlinear dependency on previous observations used as state variables in the SDDP method.

Based on the above discussion, the authors identify the following topics as promising directions for future investigation:
\begin{enumerate}
    
\item Investigation of alternative methods or SDDP approximations that can accommodate more sophisticated and realistic nonlinear forecast models. For instance, the use of regularized linear decision rules has recently demonstrated promising results \citep{Nazare}.
\item The study of non-stationary models, such as structural non-Gaussian models, to improve forecasting accuracy (we refer to \cite{bodin2020scoredrivenmodels} and \cite{hoeltgebaum2018generating}), {as well as bias correction procedures, such as the one proposed as a benchmark in Section IV.C.} 
\item The study of the causes explaining the negative recent trends through systematic testing of external data as explanatory variables and evaluation of their predictive capacity (see \cite{pina2017optimizing}).
\item The study of costs and market impacts resulting from the estimated forecast bias.
\end{enumerate}

\section{Conclusions}

In this article, we assessed the bias of Natural Inflow Energy (NIE) forecasts (the total cumulated energy potential in each subsystem) generated by the Brazilian official methodology. These bias metrics were evaluated for $k=1,...,24$ steps ahead using an out-of-sample rolling window evaluation process. Within the limitations of this study, including selected data, model hypotheses, chosen metrics, implementation methodologies, sample sizes, and benchmark models, the following conclusions can be highlighted:

\begin{enumerate}

    \item Statistically significant biases (at a significance level lower than 2.5\%) were observed from 2014 to 2024 for NIE forecasts in both the Southeast and Northeast subsystems. Figures \ref{PARp_SE}, \ref{PARp_NE}, \ref{PARpA_SE}, and \ref{PARpA_NE} provide clear evidence of the actual inadequacy of mean-reverting models (PARp and PARp-A) and the local negative trends observed in the NIE time series over the last 14 years in both Southeast and Northeast subsystems.
    
    \item Specifically, bias levels were found to be 1.28, 3.83, 5.39, and 6.73 average GW for 1-, 6-, 12-, and 24-step-ahead forecasts in the Southeast subsystem, and 0.54, 1.66, 2.32, and 3.17 average GW in the Northeast subsystem. This indicates a consistent optimistic bias in the official NIE forecasts for the two main subsystems of the Brazilian power system over the last 12 years. Furthermore, a pattern of increasing bias was identified with the number of forecast steps ahead.

    \item In general terms, the PARp-A reduces the forecast bias in comparison to the PARp model. However, no statistically significant reduction could be found for the Southeast subsystem NIE time series within the study horizon.
    
    \item Furthermore, a pattern of increasing bias was identified with the number of forecast steps ahead.
    
    \item Within the AI and machine learning models, XGBoosting and Prophet performed best in the Southeast, whereas Prophet and Chronos figured as the best and second best, respectively, in the Northeast.
    
    \item Within the test horizon, the naïve and ALTM PARp-A benchmarks exhibited cumulative forecast biases lower than or equal to 35\% of the Official PARp-A bias value in the Southeast and 11\% in the Northeast, showcasing concerning evidence that the Official methodology should be revised. 
    
    \item {Finally, the 24-months ahead cumulative bias metric for the official PARp-A model also showcases concerning bias values for both subsystems, being consistently outperformed by benchmark models. This constitutes relevant evidence that planning decisions based on the official forecast methodology should be inherently biased by an overly optimistic view of future inflows. This poses a significant concern for the sustainable utilization of water resources, which are crucial renewable energy resources of the Brazilian power system, and underscores the need for establishing new, specific, and expedited monitoring procedures for both the data and the models utilized in dispatch and planning activities.}

\end{enumerate}

\appendix  % Start the appendix

\section{Simulation procedure}

In this section, we outline the key steps of the Brazilian official methodology used to simulate future values of the NIE for all subsystems. Here, time $t$ refers to the period immediately preceding the start of the simulation horizon, following the estimation data. We adopt the same temporal framework as in Section II to generate future scenarios for $K$ steps ahead, i.e., $t+1,..., t+K$. The official methodology uses these simulated scenarios to produce probabilistic forecasts for $k=1,..., K$. As the goal of this study is to analyze the bias of the $K$ predictive distributions and detect if there is a distribution shift or not, we use the generated scenarios to calculate the $K$-step ahead expected values (based on the sample average of these scenarios) and assess the bias according to (1) and (3).

Within this context, given the estimated periodic parameters (coefficients, means, and variances), the third parameter, $\lambda_{t+k}$, is dynamically defined to ensure a non-negative time series. Specifically, $\lambda_{t+k}$ establishes a conditional lower bound for $\epsilon_{t+k}$ in the simulation procedure to maintain the positiveness of the time series. Consequently, it is a function of 1) the previous observation data, which defines the conditional expected value for the time series at time $t$, and 2) the parameters of the underlying normal distribution used to generate values for the log-normal errors. According to \cite{charbeneau1978comparison}, the three-parameter log-normal errors can be defined as follows:
\begin{align}
    \epsilon_{t+k}= e^{\xi_{t+k}} + \lambda_{t+k} \label{LogNormal_eq}\\
    \xi_{t+k}\sim N(\mu^{(\xi)}_{t+k}, \sigma^{(\xi)}_{t+k}).
\end{align}
%is the underlying normal distribution with parameters $\mu^{(\xi)}_{t}$ and $\sigma^{(\xi)}_{t}$. 

According to \cite{GTMetodologiaCPAMP2022}, for a given time step $t$ (consider the PARp model with order $p_{m_{t+k}} = 1$, for didactic purposes), the estimated model for $y_{t+k}$ follows equation \eqref{PARp_model}. By imposing $y_{t+k} > 0$, we obtain the following
\begin{align}
   \hspace{-0.15cm}y_{t+k} = \hat{\mu}_{m_{t+k}} +  \hat{\sigma}_{m_{t+k}}\hat{\phi}_1^{(m_{t+k})}\bigg(\dfrac{y_{t+k-1} - \hat{\mu}_{m_{t+k-1}}}{\hat{\sigma}_{m_{t+k-1}}}\bigg) +\notag \\\hspace{-0.15cm}\hat{\sigma}_{m_{t+k}}\epsilon_{t+k} > 0. \label{DeltaDerivation_1_EQ}
    %\hat{\sigma}_{m_{t+k}}\epsilon_{t+k} > 0. %\label{DeltaDerivation_1_EQ}
\end{align}
Thus, $\epsilon_{t+k}$ can be isolated in inequality \eqref{DeltaDerivation_1_EQ} to generate a lower bound $\lambda_{t+k}$ for its possible values as follows: 
\begin{align}
  \epsilon_{t+k} > \hat{\lambda}_{t+k}=-\dfrac{\hat{\mu}_{m_{t+k}}}{\hat{\sigma}_{m_{t+k}}} -  \hat{\phi}_1^{(m_{t+k})}\bigg(\dfrac{y_{t+k-1} - \hat{\mu}_{m_{t+k-1}}}{\hat{\sigma}_{m_{t+k-1}}}\bigg). \label{LogNormal_LowerBound_eq}
\end{align}
By construction, $\hat{\lambda}_{t+k}$ is a function of the vector of past observations, $y_{[t+k-1]}$, with observed data up to period $t+k-1$. 

To generate scenarios for $\epsilon_{t+k}$, we need to define the parameters of the underlying normal distribution,  $\mu^{(\xi)}_{t+k}$ and $\sigma^{(\xi)}_{t+k}$, based on $\hat{\lambda}_{t+k}$, $\hat{\mu}_{m_{t+k}}^{(\epsilon)}$, and $\hat{\sigma}_{m_{+k}}^{(\epsilon)}$. Note that since the parameters of $\xi_{t+k}$ will be estimated based on $\hat{\lambda}_{t+k}$, which is a function of previously observed data $y_{[t+k-1]}$, they will be different for each time, and are thus indexed by $t+k$.

First, let us obtain the mean and variance for $\epsilon_{t+k}$:
\begin{align}
    \mu_{m_{+k}}^{(\epsilon)} = &E[\epsilon_{t+k}~|~y_{[t+k-1]}] = \nonumber \\   &E\big[e^{\xi_{t+k}}~|~y_{[t+k-1]}\big] + \lambda_{t+k} \label{LogNormal_Mean_eq} \\
    (\sigma^{(\epsilon)}_{m_{t+k}})^2 = &Var\big[\epsilon_{t+k}~|~y_{[t+k-1]}]\big] = \nonumber \\ 
    &Var\big[e^{\xi_{t+k}}~|~y_{[t+k-1]}]\big].\label{LogNormal_Variance_eq}
\end{align}
Expression \eqref{LogNormal_Mean_eq} 
provides the expected value of a two-parameter log-normal random variable shifted by $\lambda_{t+k}$, while \eqref{LogNormal_Variance_eq} corresponds to the variance of a standard two-parameter log-normal distribution. Given that $\xi_{t+k}$ is normally distributed with parameters $\mu^{(\xi)}_{t+k}$ and $\sigma^{(\xi)}_{{t+k}}$, it follows that 
\eqref{LogNormal_Mean_eq} and \eqref{LogNormal_Variance_eq} can be rewritten as: 
\begin{align}
    \mu_{m_{1+k}}^{(\epsilon)} &= e^{\mu^{(\xi)}_{t+k} + \dfrac{(\sigma^{(\xi)}_{t+k})^2} {2}} + \lambda_{t+k} \label{LogNormal_Mean_eq2} \\
    (\sigma^{(\epsilon)}_{m_{t+k}})^2 &= e^{2\big(\mu^{(\xi)}_{t+k} + \big(\sigma^{(\xi)}_{t+k}\big)^2\big)} - e^{2\mu^{(\xi)}_{t+k} + \big(\sigma^{(\xi)}_{t+k}\big)^2}. \label{LogNormal_Variance_eq2}
\end{align}

To simplify notation we follow \cite{charbeneau1978comparison} and \cite{de2008comparison} and set $ ln(\theta_{t+k}) = \big(\sigma_{t+k}^{(\xi)}\big)^2$. Thus, isolating $\mu^{(\xi)}_{t+k}$ in \eqref{LogNormal_Variance_eq2} we find:
\begin{align}
    \mu^{(\xi)}_{t+k} = \frac{1}{2}ln\bigg(\frac{({\sigma}^{(\epsilon)}_{m_{t+k}})^2}{\theta_{t+k}^2 - \theta_{t+k}} \bigg). \label{Normal_ExpectedValue}
\end{align}
The previous expressions establish the relationship between a normal distribution $\xi_{t+k}$ and the three-parameter log-normal distribution $\epsilon_{t+k}$. To estimate the normal distribution parameters, recall that the official methodology imposes $\mu_{m_{t+k}}^{(\epsilon)}=0$ and $\sigma^{(\epsilon)}_{m_{t+k}} = \hat{\sigma}^{(\epsilon)}_{m_{t+k}}$. By substituting \eqref{Normal_ExpectedValue} into \eqref{LogNormal_Mean_eq2}, and using the two moments to replace the left-hand sides of \eqref{LogNormal_Mean_eq2} and \eqref{LogNormal_Variance_eq2}, we obtain:
\begin{align}
    \hat{\theta}_{t+k} = 1 + \frac{(\hat{\sigma}^{(\epsilon)}_{m_{t+k}})^2}{\hat{\lambda}_{t+k}^2} \label{thetavalue}
\end{align}
After obtaining $\hat{\theta}_{t+k}$, we can finally calculate $\hat{\mu}_{t+k}^{(\xi)}$ and $\hat{\sigma}_{t+k}^{(\xi)}$ based on \eqref{Normal_ExpectedValue} and the definition of $\theta_{t+k}$ as follows:
\begin{align}
    \hat{\mu}^{(\xi)}_{t+k} = \frac{1}{2}ln\bigg(\frac{({\hat{\sigma}}^{(\epsilon)}_{m_{t+k}})^2}{\hat{\theta}_{t+k}^2 - \hat{\theta}_{t+k}} \bigg)
\end{align}
\begin{align}
    \big(\hat{\sigma}_{t+k}^{(\xi)}\big)^2 = ln(\hat{\theta}_{t+k}). \label{sigmavalue}
\end{align}

%Although in this work we focus on analyzing the forecast bias for the SE and NE subsystems, the Brazilian energy system is actually modeled by also considering the South and North subsystems. Therefore, as discussed in \cite{jardim2001stochastic}, after fitting the PARp or PARp-A model for each subsystem, the covariance matrix, $\hat{U}_{m}$, can be estimated from the residuals. Notice that, similarly to the estimated standard deviation $\hat{\sigma}^{(\epsilon)}_{m}$, there will be one such covariance matrix $\hat{U}_{m}$ for each month $m$. Then, we can obtain spatially correlated residuals as:
%\begin{align}
%    \hat{\eta}_{m_t} = \hat{B}_{m_t}a
%\end{align}
%where $a$ is a standard normal variable and $\hat{B}_{m_t}$ is derived from the Cholesky decomposition of $\hat{U}_{m_t}$.

The procedure used to generate $|\Omega|$ synthetic scenarios for the NIE $K$
 steps ahead is an integral part of the official methodology employed to simulate the system's operation \citep{pereira1984stochastic, pereira1991multi}. Within this framework, the simulation step involves a joint probabilistic forecast based on synthetically generated scenarios for all subsystems 
$s\in S$. This procedure is summarized in the sequel: 

\begin{enumerate}

\item Estimate the parameters of each periodic model $m\in M$ and subsystem $s$.% (PARp or PARp-A), i.e., ${\phi}^{(m_t)}_1$, $\dots$, ${\phi}^{(m_t)}_{p_{m_t}}$ and $\psi^{(m_t)}$, ${U}_{m}$, and ${B}_{m}$,

    \item For $k=1,\dots,K$ and $\omega \in \Omega$:
    \begin{enumerate}
        \item Sample a $|S|$-dimensional vector $\boldsymbol{a}_{t+k,\omega}\sim N(\boldsymbol{0},\boldsymbol{I})$ and obtain spatially-correlated residuals $\boldsymbol{\eta}_{t+k} = \hat{\boldsymbol{B}}_{m_{t+k}}\boldsymbol{a}_{t+k,\omega}$.
        \item Evaluate ${{\hat{\lambda}}}_{t+k,\omega}$ using \eqref{LogNormal_LowerBound_eq} and then obtain ${\hat{\mu}}^{(\xi)}_{t+k,\omega}$ and ${\hat{\sigma}}^{(\xi)}_{t+k,\omega}$ using \eqref{thetavalue}--\eqref{sigmavalue} for each subsystem $s\in S$. Store the simulated values in $|S|$-dimensional vectors.
        \item Generate a sample for the ${\xi}_{t+k,\omega} = {\hat{\sigma}}^{(\xi)}_{t+k,\omega}{\eta}_{t+k,\omega} +{\hat{\mu}}^{(\xi)}_{t+k,\omega}$ for each subsystem $s\in S$.
        \item Obtain a sample for $\epsilon_{t+k,\omega}$ using $\hat{\epsilon}_{t+k,\omega}= e^{\hat{\xi}_{t+k,\omega}} + {\hat{\lambda}}_{t+k,\omega}$ for each subsystem $s\in S$.
        \item Finally, obtain a sample of $\hat{y}_{t+k|y_{[t]}}^{(\omega)}$ using \eqref{PARp_model} or \eqref{PARpA_model} for each subsystem $s\in S$. From this sample, a probabilistic forecast ($\hat{F}_{t+k|t})$ or any point forecast $(\hat{y}_{t+k|y_{[t]}})$, based on the mean or quantiles, can be derived.  
    \end{enumerate} 
\end{enumerate}

We conclude with a final remark regarding some aspects of the simulations obtained from the PARp and PARp-A models. One has to keep in mind that higher-order moments, such as skewness and kurtosis of the log-normal, will depend on the parameters of the underlying normal distribution.

\bibliography{Bibliography}

\end{document}